\documentclass[prd,showpacs,preprintnumbers,amsmath,amssymb]{revtex4}
\usepackage{graphics}
\usepackage{epsfig}
\usepackage{dcolumn}% Align table columns on decimal point
\usepackage{bm}% bold math

\begin{document}
%\begin{CJK}{GBK}{song}
\title{$Y(4143)$ is probably a molecular partner of $Y(3930)$}
\author{Xiang Liu$^{1,2}$}\email{xiangliu@pku.edu.cn}
\author{Shi-Lin Zhu$^1$\footnote{Corresponding author}}
\email{zhusl@phy.pku.edu.cn} \affiliation{$^1$Department of
Physics, Peking University, Beijing 100871, China\\$^2$Centro de
F\'{i}sica Computacional, Departamento de F\'{i}sica, Universidade
de Coimbra, P-3004-516 Coimbra, Portugal}

\date{\today}% It is always \today, today,

\begin{abstract}

After discussing the various possible interpretations of the
$Y(4143)$ signal observed by the CDF collaboration in the $J/\psi
\phi$ mode, we tend to conclude that $Y(4143)$ is probably a
$D_s^\ast {\bar D}_s^\ast$ molecular state with $J^{PC}=0^{++}$ or
$2^{++}$ while $Y(3930)$ is its $D^\ast {\bar D}^\ast$ molecular
partner as predicted in our previous work \cite{we}. Both the
hidden-charm and open charm two-body decays occur through the
rescattering of the vector components within the molecular states
while the three- and four-body open charm decay modes are
forbidden kinematically. Hence their widths are narrow naturally.
CDF, Babar and Belle collaborations may have discovered heavy
molecular states already. We urge experimentalists to measure
their quantum numbers and explore their radiative decay modes in
the future.

\end{abstract}

\pacs{12.39.Pn, 12.40.Yx, 13.75.Lb}

\maketitle

%%%%%%%%%%%%%%%%%%%%%%%%%%%%%%%%%%%%%%%%%%%
\section{Introduction}\label{sec1}
%%%%%%%%%%%%%%%%%%%%%%%%%%%%%%%%%%%%%%%%%%%

The charmonium system has been a rich gold mine, which continues to
deepen our knowledge of strong interaction. Especially starting from
2003 many new charmonium or charmonium-like states were discovered
experimentally, some of which do not easily fit into the
conventional quark model picture. Let's list these new states:
$X(3872)$, $X(3940)$/$Y(3930)$/$Z(3930)$, $Y(4260)$, $Z(4430)$ etc.
There have been many surprises and unexpected in this active field.
The most recent one came from the CDF Collaboration
\cite{cdf-y4143}.

In the exclusive $B^+\to J/\psi \phi K^+$ decays, the CDF
collaboration observed a narrow structure $Y(4143)$ near the $J/\psi
\phi$ threshold ($m_{J/\psi}+m_\phi=4.117$ GeV) with a statistical
significance of $3.8 \sigma$. The mass and width of this signal are
measured to be $4143\pm 2.9\pm 1.2$ MeV and $11.7^{+8.3}_{-5.0}\pm
3.7$ MeV \cite{cdf-y4143}. For comparison, $Y(4143)$ is very similar
to the charmonium-like state $Y(3930)$ which was observed by both
Belle and Babar collaborations near the $J/\psi \omega$ threshold
($m_{J/\psi}+m_\omega=3.88$ GeV) \cite{belle-y3930,babar-y3930}. The
mass and width of Y(3930) are $3914.6^{+3.8}_{-3.4}\pm 2.0$ MeV and
$34^{+12}_{-8}\pm 5$ MeV \cite{babar-y3930}.

Clearly the C-parity and G-parity of $Y(4143)$ are even. Around
this mass region, $Y(4143)$ may naively be speculated to be an
excited charmonium state. We list some possibilities below:
\begin{enumerate}
\item{ the second radial excitation of the P-wave charmonium:
$\chi_{cJ}^{\prime\prime}$\\

Sometimes either $X(3872)$ or $X(3930)$ is speculated to be
$\chi_{c1}^{\prime}$. Then $Y(4143)$ seems plausible as a candidate
of $\chi_{cJ}^{\prime\prime}$. }

\item{$\eta_c(3S)$\\

As $\eta_c(3S)$, its mass seems too high compared with $\psi(3S)$
at 4040 MeV. }

\item{Other higher lying states such as $2 ^1D_2, 1 ^3F_{2,3,4}$
etc}.
\end{enumerate}
However as emphasized by CDF collaboration \cite{cdf-y4143},
$Y(4143)$ lies well above the open charm decay threshold.
Charmoniums with this mass would decay into an open charm pair
dominantly. The branching fraction of its hidden charm decay mode
$J/\psi \phi$ is expected to be tiny. Thus, both the narrow width
of $Y(4143)$ and its large hidden charm decay pattern disfavor the
conventional charmonium interpretation.

One may be tempted to try the non-conventional assignment of
$Y(4143)$ as a tetraquark state. In fact, the $J^P=0^+, 1^+$
$c\bar c s \bar s$ tetraquark states were predicted to be around
4.1 GeV in the simple chromo-magnetic interaction quark model
while their $J^P=2^+$ partner seems unbound \cite{tetra}. However
tetraquarks will fall apart into a pair of charmed mesons very
easily. As a tetraquark, the width of $Y(4143)$ would be around
several hundred MeV instead of 12 MeV as observed by CDF.

The $B$ meson decay process does not provide a glue-rich environment
as in the Upsilon/charmonium annihilation. So the possibility of
$Y(4143)$ being a glueball is small.

Recall that $Y(4260)$ is proposed as the $1^{--}$ hybrid
charmonium state \cite{hybrid1,hybrid2,hybrid3}. All current
experimental information is compatible with this assumption
\cite{myreview}. The only slightly worrisome issue is the leptonic
width $\Gamma [Y(4260)\to e^+e^-]\approx 0.8 $ keV if one assumes
$B[Y(4260)\to J/\psi \pi^+\pi^-] \approx 1\%$. Such a leptonic
width naively seems a little large for a vector hybrid.
Fortunately no dynamical selection rule has been discovered to
suppress the leptonic width of a hybrid state up to now.

Within this context, $Y(4143)$ could be an exotic charmonium with
$J^P=1^{-+}$, which is a partner of $Y(4260)$ within the same
$J^P=1^-$ hybrid charmonium family. One may challenge this
assignment with the following three facts: (1) the narrow width of
$Y(4143)$ which is much smaller than that of $Y(4260)$; (2) the
different decay patterns of $Y(4143)$ and $Y(4260)$. $Y(4260)$ is
observed in $J/\psi \pi^+\pi^-$, $J/\psi K^+ K^-$ modes while
$Y(4143)$ only in $J/\psi \phi$; (3) the similarity between
$Y(4143)$ and $Y(3930)$. One does not expect two exotic $1^{-+}$
states lying so closely.

In fact, the extreme similarity between $Y(4143)$ and $Y(3930)$
suggests that they be partner molecular states belonging to the
same representation as predicted in our previous work \cite{we},
where we have performed a systematic study of the possible
molecular states composed of a pair of heavy mesons in the
framework of the meson exchange model. $Y(4143)$ is a $D_s^\ast
{\bar D}_s^\ast$ molecular state while $Y(3930)$ is its $D^\ast
{\bar D}^\ast$ molecular partner. Their hidden charm decays occur
through the rescattering mechanism \cite{rescattering}. We present
more details in Sections \ref{sec2}-\ref{sec3}.

%%%%%%%%%%%%%%%%%%%%%%%%%%%%%%%%%%%%%%%%%%%%%%
\section{$Y(4143)$ and $Y(3930)$ as heavy molecular candidates}\label{sec2}
%%%%%%%%%%%%%%%%%%%%%%%%%%%%%%%%%%%%%%%%%%%%%%

Since $Y(4143)$ and $Y(3930)$ were observed in the $J/\psi\phi$
and $J/\psi\omega$ modes, these two states are isoscalars. The
isoscalar heavy molecular states in the vector-vector channel are
denoted as $\Phi_8^{\ast\ast 0}$ and $\Phi_{s1}^{\ast\ast 0}$
respectively in our previous work \cite{we}. From now on we
identify $\Phi_8^{\ast\ast 0}$ as $Y(3930)$ and
$\Phi_{s1}^{\ast\ast 0}$ as $Y(4143)$. Their flavor wave functions
are
\begin{eqnarray*}
|Y(3930)\rangle&=&\frac{1}{\sqrt{2}}[|D^{*0}\bar{D}^{*0}\rangle+|D^{*+}D^{*-}\rangle],\\
|Y(4143)\rangle&=&|D_s^{*+}D_s^{*-}\rangle.
\end{eqnarray*}

We used the following effective Lagrangians to derive the meson
exchange potentials of $Y(4143)$ and $Y(3930)$:
\begin{eqnarray} \nonumber
\mathcal{L}_{\mathcal{D^*D^*}\mathbb{P}}&=&\frac{1}{2}g_{\mathcal{D^*D^*}\mathbb{P}}
\varepsilon_{\mu\nu\alpha\beta}\left(D_{a}^{*\mu}\partial^\alpha
\mathcal{D}_{b}^{*\beta\dag}
-\mathcal{D}_{b}^{*\beta\dag}\partial^\alpha
 \mathcal{D}_{a}^{*\mu}\right)\partial^\nu \mathbb{P}_{ab},
\label{la-3}\\ \nonumber \mathcal{L}_{\mathcal{D^*D^*}\mathbb{V}
}&=&ig_{\mathcal{D^*D^*}\mathbb{V}}\left(\mathcal{D}_{a}^{*\nu\dag}\partial^\mu
\mathcal{D}_{\nu,b}^{*}-\mathcal{D}_{\nu b}^{*}\partial^\mu
\mathcal{D}_{a}^{*\nu\dag}\right)(\mathbb{V}_\mu)_{ab}
+4if_{\mathcal{D^*D^*}\mathbb{V}}D_{\mu a}^{*\dag} D_{\nu
b}^{*}(\partial^\mu \mathbb{V}^\nu -\partial^\nu
\mathbb{V}^\mu)_{ab}, \label{la-4}\\ \nonumber
\mathcal{L}_{\mathcal{D^*D^*}\sigma
}&=&2\,m_{\mathcal{D}^*}\,g_\sigma \mathcal{D}_{a}^{*\alpha}
\mathcal{D}_{\alpha a}^{*\dag}\sigma,\label{la-5},
\end{eqnarray}
where $\mathcal{D^{(*)}}$=(($\bar{D}^{0})^{(*)}$, $(D^{-})^{(*)}$,
$(D_{s}^{-})^{(*)}$). $\mathbb{P}$ and $\mathbb{V}$ are the octet
pseudoscalar and nonet vector meson matrices respectively.

In general, the possible quantum numbers of the S-wave
vector-vector system are $J^{P}=0^+, 1^+, 2^+$. However for the
neutral $D^\ast {\bar D}^\ast$ system with $C=+1$, we can have
$J^{P}=0^+$ and $2^+$ only since $C=(-)^{L+S}$ and $J=S$ with
$L=0$. The exchanged mesons include the pseudoscalar, vector and
$\sigma$ mesons. Details of the derivation of the exchange
potential can be found in Ref. \cite{we}. We collect the resulting
expressions below:
\begin{eqnarray}
\mathcal{V}(r)_{Y(4143)}^{[J]}&=&\frac{1}{6}g_{\mathcal{D^*D^*}\mathbb{P}}^2\,\mathcal{A}[J]Z[\Lambda,m_{\eta},r]
-
\Bigg[\frac{g_{\mathcal{D^*D^*}\mathbb{V}}^2}{4m_{\mathcal{D}_s^*}^2m_\phi^2}
\,\mathcal{C}[J]\,X[\Lambda,m_\phi,r]\nonumber\\&&+g_{\mathcal{D^*D^*}
\mathbb{V}}^2\,\mathcal{C}[J]\,Y[\Lambda,m_\phi,r]
+\frac{4f_{\mathcal{D^*D^*}\mathbb{V}}^2}{m_{\mathcal{D}_s^*}^2}\,\mathcal{B}[J]\,
Z[\Lambda,m_\phi,r]\Bigg],\label{vv-4}
\end{eqnarray}
\begin{eqnarray}
\mathcal{V}(r)^{[J]}_{Y(3930)}&=&
g_{\mathcal{D^*D^*}\mathbb{P}}^2\,\mathcal{A}[J]\,\left[\frac{3}{8}Z[\Lambda,m_{\pi},r]
+\frac{Z[\Lambda,m_{\eta},r]}{24}\right]-{g_\sigma^2}
\,\mathcal{C}[J]\,Y[\Lambda,m_\sigma,r]\nonumber\\&&- \frac{3}{2}
\Bigg[\frac{g_{\mathcal{D^*D^*}\mathbb{V}}^2}{4m_{\mathcal{D}^*}^2m_\rho^2}
\,\mathcal{C}[J]\,X[\Lambda,m_\rho,r]+g_{\mathcal{D^*D^*}
\mathbb{V}}^2\,\mathcal{C}[J]\,Y[\Lambda,m_\rho,r]
+\frac{4f_{\mathcal{D^*D^*}\mathbb{V}}^2}{m_{\mathcal{D}^*}^2}\,\mathcal{B}[J]\,Z[\Lambda,m_\rho,r]\Bigg]\nonumber\\&&
- \frac{1}{2}
\Bigg[\frac{g_{\mathcal{D^*D^*}\mathbb{V}}^2}{4m_{\mathcal{D}^*}^2m_\omega^2}
\,\mathcal{C}[J]\,X[\Lambda,m_\omega,r]+g_{\mathcal{D^*D^*}
\mathbb{V}}^2\,\mathcal{C}[J]\,Y[\Lambda,m_\omega,r]
+\frac{4f_{\mathcal{D^*D^*}\mathbb{V}}^2}{m_{\mathcal{D}^*}^2}\,\mathcal{B}[J]\,Z[\Lambda,m_\omega,r]\Bigg]\nonumber\\\label{vv-3}
\end{eqnarray}
with $Y[\Lambda,m,r]=\frac{1}{4\pi r}\left(e^{-mr}-e^{-\Lambda
r}\right)-\frac{\xi^2}{8\pi \Lambda}e^{-\Lambda r}$,
$Z[\Lambda,m,r]=-\frac{1}{r^2}\frac{\partial}{\partial r}
\left(r^2\frac{\partial}{\partial r}\right)Y[\Lambda,m,r]$,
$X[\Lambda,m,r]=\left[-\frac{1}{r^2}\frac{\partial}{\partial
r}\left(r^2\frac{\partial}{\partial
r}\right)+m^2\right]Z[\Lambda,m,r]$, where
$\xi=\sqrt{\Lambda^2-m^2}$. $\Lambda$ arises from the monopole
form factor. The coefficients $\mathcal{A}(J)={2\over 3}$,
$\mathcal{B}(J)={4\over 3}$ and $\mathcal{C}(J)=1$ for J=0 while
$\mathcal{A}(J)=-{1\over 3}$, $\mathcal{B}(J)=-{2\over 3}$ and
$\mathcal{C}(J)=1$ for J=2.

\begin{figure}[htb]\begin{center}
\begin{tabular}{cc}
\scalebox{0.7}{\includegraphics{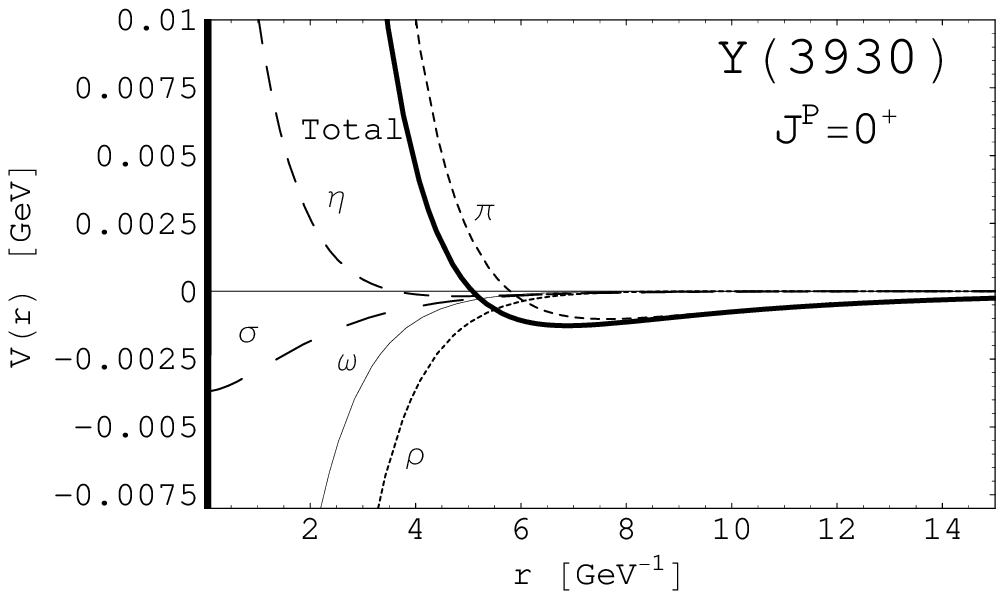}}&
\scalebox{0.7}{\includegraphics{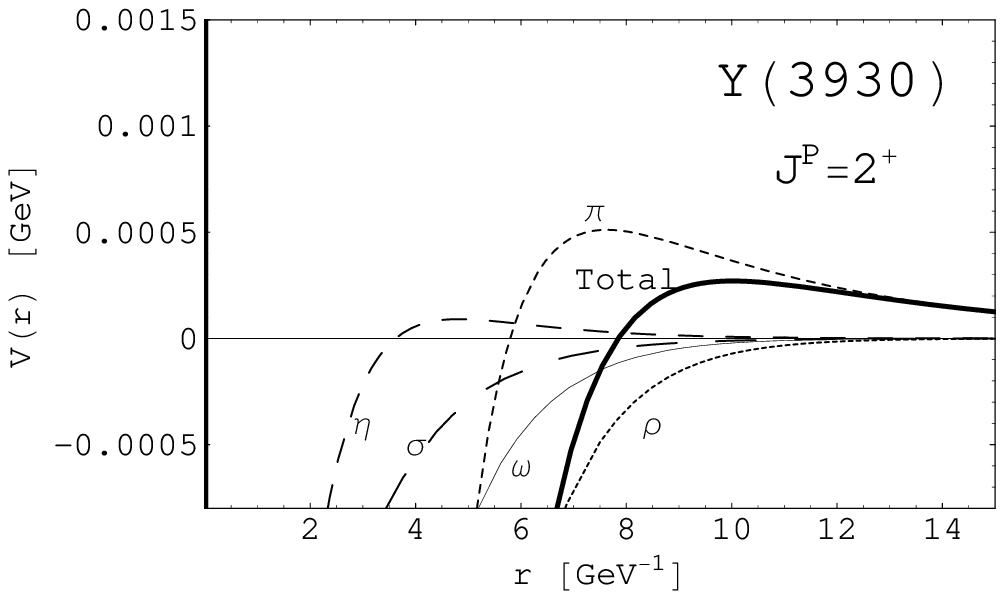}}\\(a)&(b)\\
\end{tabular}
\caption{The shape of the exchange potential of $Y(3930)$. The
thick solid line is the total effective potential. Here we take
the cutoff $\Lambda=1$ GeV. \label{VV-potential-3}}\end{center}
\end{figure}

\begin{figure}[htb]\begin{center}
\begin{tabular}{cc}
\scalebox{0.7}{\includegraphics{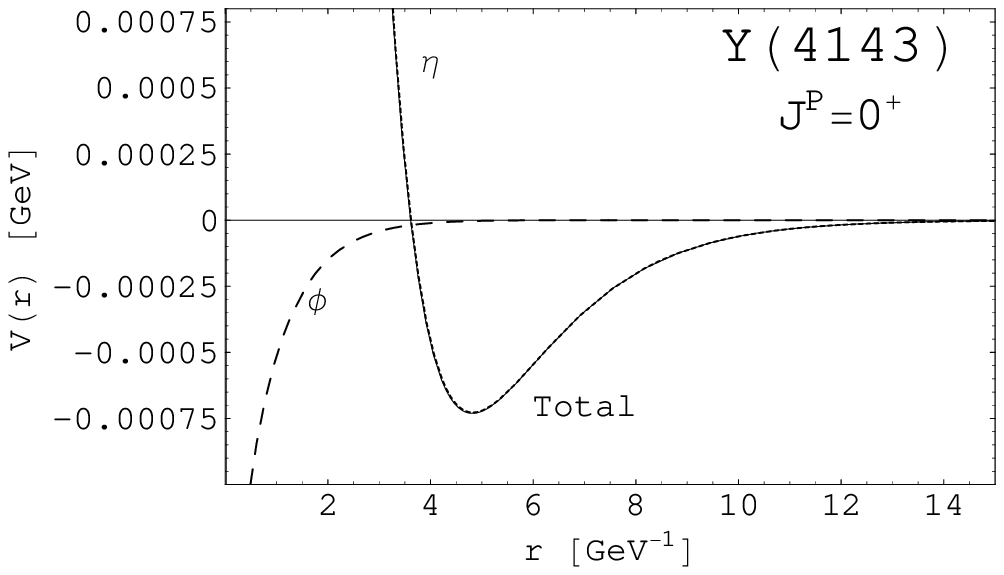}}&
\scalebox{0.7}{\includegraphics{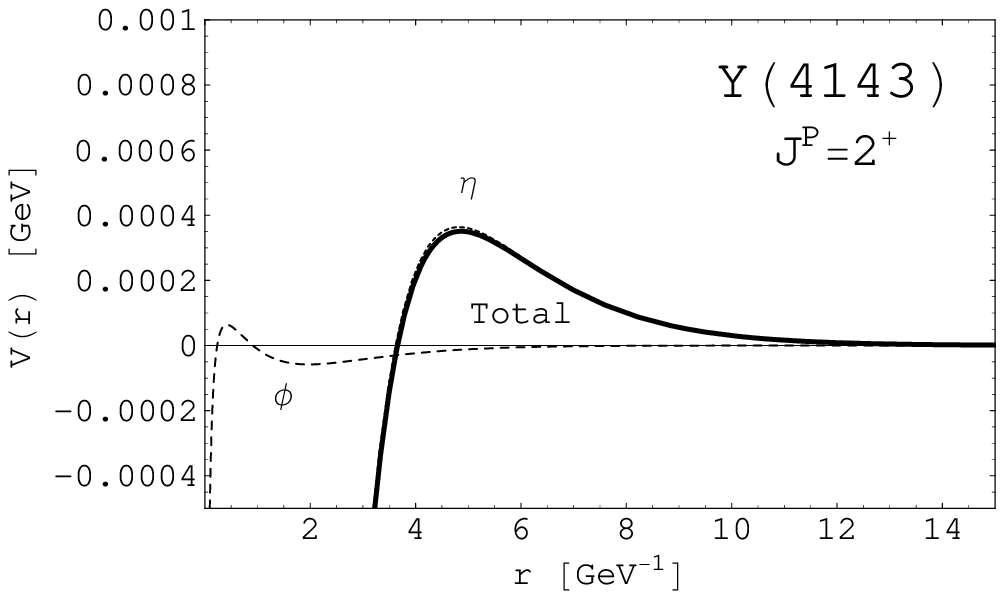}}\\(a)&(b)\\
\end{tabular}
\caption{The variation of the effective potential of $Y(4143)$
with $r$. The thick solid line is the total effective potential.
Here we take the cutoff $\Lambda=1$
GeV.\label{VV-potential-4}}\end{center}
\end{figure}

\begin{figure}[htb]\begin{center}
\begin{tabular}{cc}
\scalebox{0.7}{\includegraphics{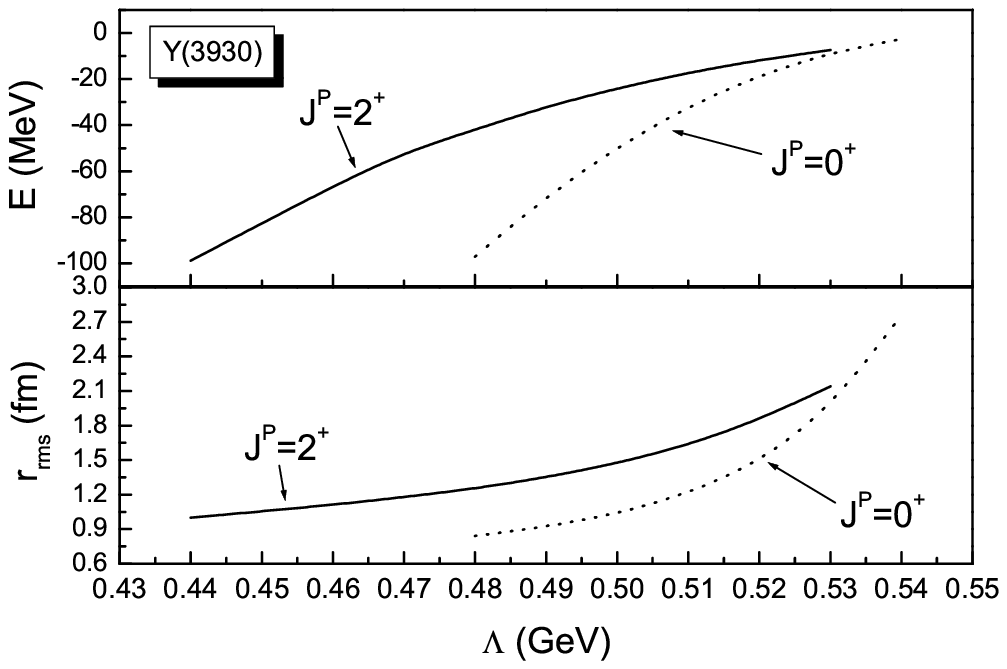}}&
\scalebox{0.7}{\includegraphics{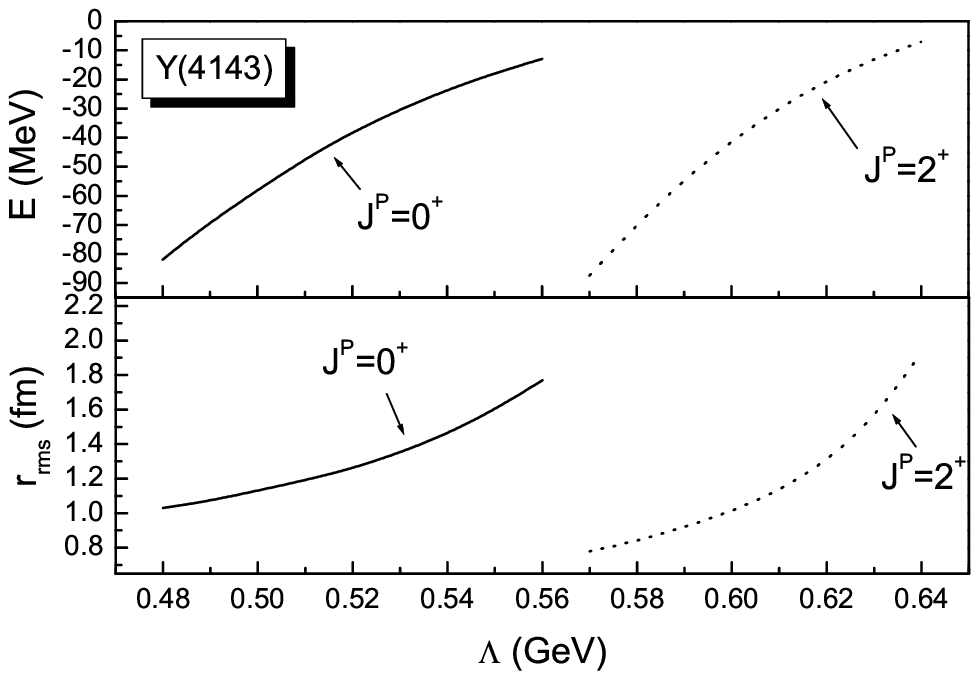}}
\end{tabular}
\caption{The variation of the binding energy $E$ and $r_{rms}$ with
the cutoff $\Lambda$. \label{diagram}}\end{center}
\end{figure}

The effective potentials of $Y(3930)$ with $J^P=0^+, 2^+$ are
presented in Fig. \ref{VV-potential-3}, where the exchanged mesons
include $\pi$, $\eta$, $\rho$, $\omega$ and $\sigma$. For
$Y(4143)$, its effective potential is shown in Fig.
\ref{VV-potential-4}, where both the $\eta$ and $\phi$ meson
exchange is allowed. Clearly there exists quite strong attraction
around 1 fm for both $Y(3930)$ and $Y(4143)$ with $J^P=0^+$. Then
we solved the Schr\"{o}dinger equation to find the binding energy
for the $Y(3930)$ and $Y(4140)$ systems. Some typical numerical
results are collected in Table \ref{DsDs} and Fig. \ref{diagram}.
We found molecular solutions for $Y(3930)$ and $Y(4140)$ with
$J^P=0^+, 2^+$. The first line in Table \ref{DsDs} is added after
CDF released their data, which corresponds to the central mass of
Y(3930) and Y(4143).


\begin{center}
\begin{ruledtabular}
\begin{table}[htb]
\begin{tabular}{c||ccc|cccccc}

&\multicolumn{3}{c}{$J^P=0^+$}&\multicolumn{3}{c}{$J^P=2^+$}\\\hline

State&$\Lambda$ (GeV)&$E$ (MeV)& $r_{\mathrm{rms}}$ (fm)&$\Lambda$
(GeV)&$E$ (MeV)& $r_{\mathrm{\mathrm{rms}}}$ (fm)\\\hline

$Y(3930)$

&0.48&-97.1&0.84&0.44&-98.9&1.00\\
&0.49&-71.2&0.92&0.46&-66.4&1.11\\
&0.50&-49.5&1.03&0.48&-41.9&1.25\\
&0.51&-31.92&1.21&0.50&-24.0&1.47\\
&0.52&-18.32&1.48&0.51&-17.23&1.63\\
&0.53&-8.61&1.95&0.52&-11.71&1.85\\
&0.54&-2.57&2.76&0.53&-7.36&2.14\\\hline

$Y(4143)$

&0.57&-87.4&0.78&0.48&-81.9&1.03\\
&0.58&-69.9&0.84&0.49&-68.9&1.07\\
&0.59&-54.5&0.92&0.51&-47.2&1.19\\
&0.60&-41.2&1.01&0.52&-38.2&1.26\\
&0.61&-29.9&1.13&0.53&-30.4&1.35\\
&0.62&-20.46&1.30&0.54&-23.64&1.46\\
&0.63&-12.91&1.55&0.55&-17.87&1.60\\
&0.64&-7.13&1.93&0.56&-13.01&1.77\\

            \end{tabular}
\caption{Some typical numerical results for the $Y(3930)$ and
$Y(4143)$ systems.\label{DsDs}}
\end{table}
\end{ruledtabular}
\end{center}

%%%%%%%%%%%%%%%%%%%%%%%%%%%%%%%%%%%%%%%%%%%%%%%%%%%%%%%%%
\section{Characteristic Decay Patterns of $Y(4143)$ and $Y(3930)$}\label{sec3}
%%%%%%%%%%%%%%%%%%%%%%%%%%%%%%%%%%%%%%%%%%%%%%%%%%%%%%%%%

Now we discuss the possible decay modes of these molecular states.
\begin{enumerate}
\item{Hidden charm two-body decay\\
This is the discovery mode of $Y(3930)$ and $Y(4143)$. This S-wave
decay occurs through the rescattering mechanism
\cite{rescattering}, which is shown in Fig. \ref{re3} (a). The
branching ratio of the hidden charm decay is strongly suppressed
for the conventional excited charmonium around 4 GeV. However,
this mode is certainly not suppressed for heavy molecular states.
Sometimes, they become one of the dominant decay modes.

\begin{figure}
\begin{tabular}{ccc}
\scalebox{0.5}{\includegraphics{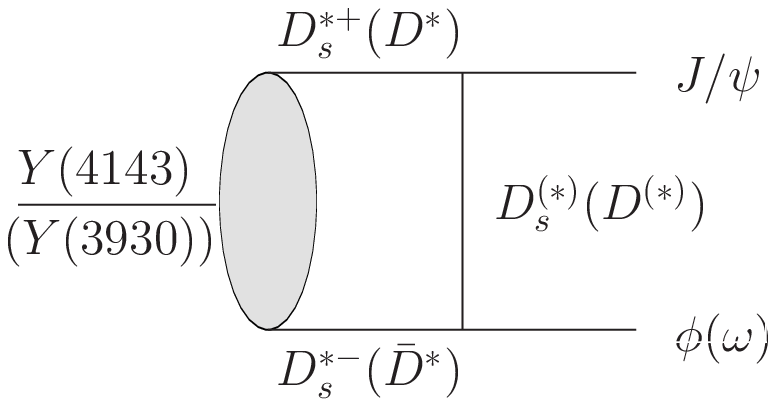}}&
\scalebox{0.5}{\includegraphics{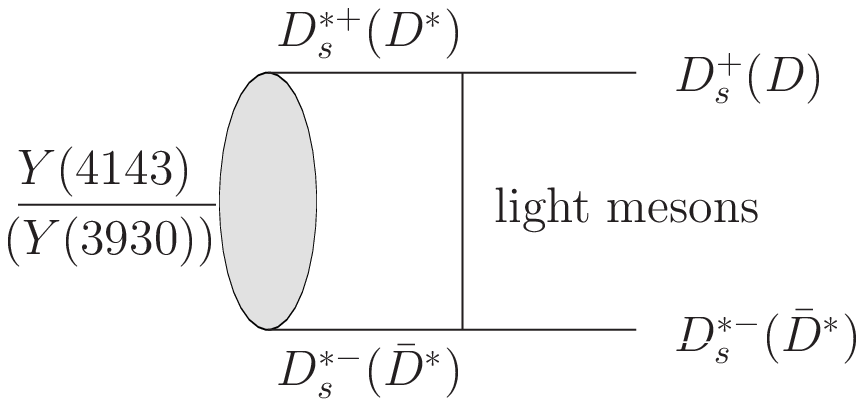}}&
\scalebox{0.5}{\includegraphics{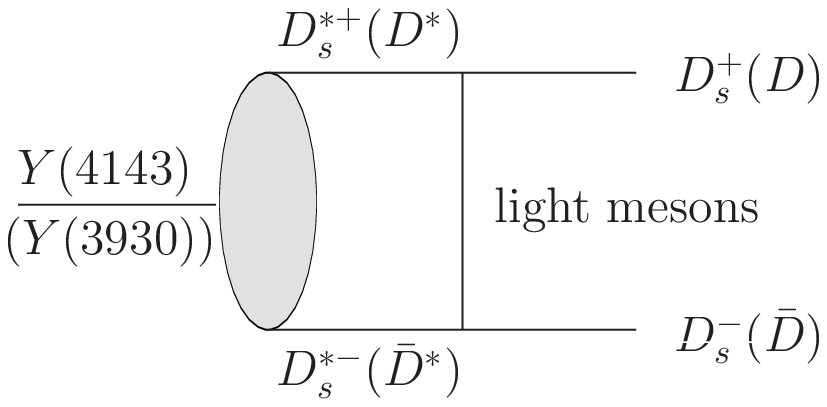}}\\
(a)&(b)&(c)\\\end{tabular} \caption{Rescattering diagram for the
hidden and open charm decays of $Y(4143)$ and $Y(3930)$
\label{re3}.}
\end{figure}
}

\item{Open charm two-body decay\\

At the first sight, one may worry whether the open-charm decays of
$Y(3930)$ and $Y(4143)$ overwhelm the hidden decay as in the
excited charmonium decays. However, this is not the case for
molecular states. In fact, the kinematically allowed modes are
$D_s^{+}D_s^{*-}+h.c./D\bar{D}^*+h.c.$ and
$D_s^{+}D_s^{-}+h.c./D\bar{D}+h.c.$. These decays can only happen
via the exchange of a light meson as shown in Fig. \ref{re3} (b)
and (c). In other words, the open-charm decay widths are
comparable to the hidden-charm width. There are some simple
selection rules from the parity and angular momentum conservation.
For example, the $J^P=0^+$ state does not decay into $ D\bar
D^\ast (D^+_s D_s^{\ast -})$. The $J^P=2^+$ state decays into $
D\bar D^\ast (D^+_s D_s^{\ast -})$ or $ D \bar D (D^+_s D_s^{ -})$
through D-wave. Such a D-wave decay width should be much smaller
than that for the above hidden charm S-wave decay mode.

}

\item{Open charm three-body and four-body decays\\
Another potentially important decay mode is that one or two
components of the molecular state decays into a $D\pi$ or $DK$
pair. The possible three- and four-body decay modes of $Y(3930)$
are $\bar D^\ast D\pi$, $\bar D D \pi\pi$. For $Y(4143)$, the
possible modes are ${\bar D}_s^\ast DK$, $\bar D K DK$. However
all these three- and four-body decay modes are kinematically
forbidden! The isospin violating modes ${\bar D}_s^\ast D_s \pi^0$
and ${\bar D}_s \pi^0 D_s \pi^0$ are also kinematically forbidden
for $Y(4143)$.

}

\item{Radiative decay\\
One or both vector mesons can easily decay into the $D\gamma$ or
$D_s\gamma$ final states. The typical radiative modes of $Y(3930)$
are ${\bar D}^\ast D\gamma$, ${\bar D}\gamma D\gamma$, ${\bar
D}\pi D\gamma$. For $Y(4143)$ the radiative modes are ${\bar
D}_s^\ast D_s \gamma$, ${\bar D}_s\gamma D_s \pi^0$, ${\bar
D}_s\gamma D_s \gamma$. The radiative decay width and the line
shape of the photon spectrum are very interesting, which will be
pursued in a subsequent work.

}

\item{Semi-leptonic and Non-leptonic decay\\
The semi-leptonic and non-leptonic decays via one component of the
molecular state also contain useful information of its inner
structure.

}

\end{enumerate}

%%%%%%%%%%%%%%%%%%%%%%%%%%%%%%%%%%%%%%%%%%%%%%%%%%%%%%%%%
\section{Conclusion}\label{sec4}
%%%%%%%%%%%%%%%%%%%%%%%%%%%%%%%%%%%%%%%%%%%%%%%%%%%%%%%%%

In this paper we have discussed the various assignment of the
$Y(4143)$ signal discovered by the CDF collaboration. With the
extreme similarity between $Y(3930)$ and $Y(4143)$ in mind, we
conclude that both of them are very good molecular states composed
of a pair of vector charm mesons, which was predicted in the meson
exchange model \cite{we}. $Y(3930)$ and $Y(4143)$ are molecular
partners with $J^{PC}=0^{++}$ or $2^{++}$. Such a classification
neatly explains the narrow width of $Y(4143)$. The hidden charm
discovery mode $J/\psi\phi$ naturally becomes one of the dominant
decay modes since the kinematically allowed open-charm decays also
occur through the same rescattering mechanism. Hence the width of
the hidden-charm and open charm decay widths are comparable.
Especially for the case of $J^{PC}=2^{++}$ the hidden charm decay
is the dominant decay mode since the two-body open charm decay
occurs through D-wave.

In short summary, CDF, Babar and Belle collaborations may have
discovered heavy molecular states already. We urge our
experimental colleagues to confirm Y(4143) and measure its $J^P$
after accumulating more $J\psi\phi$ data, or even in the radiative
decays.

In fact the above interpretation helps us to clarify these new
charmonium states observed in the past six years:
\begin{enumerate}
\item{$Z(3930)$ is a $2^{++}$ charmonium
$\chi_{c2}^{\prime\prime}$;}

\item{$X(3940)$ may be $\eta_c(3S)$;}

\item{Both $Y(4143)$ and $Y(3930)$ are probably molecular states;}

\item{$Y(4260)$ is a good candidate of the hybrid charmonium;}

\item{$X(3872)$ may be $\chi_{c1}^\prime$ or a $\bar DD^\ast$
molecular state;}

\item{$Z^+(4430)$ could be a molecular candidate however its
existence needs further confirmation.}

\end{enumerate}
%%%%%%%%%%%%%%%%%%%%%%%%%%%%%%%%
\section*{Acknowledgments}
%%%%%%%%%%%%%%%%%%%%%%%%%%%%%%%%

This project was supported by National Natural Science Foundation
of China under Grants 10625521, 10721063 and 10705001. One of us
(X.L.) is support in part by the \emph{Funda\c{c}\~{a}o para a
Ci\^{e}ncia e a Tecnologia} \/of the \emph{Minist\'{e}rio da
Ci\^{e}ncia, Tecnologia e Ensino Superior} \/of Portugal, under
contract SFRH/BPD/34819/2007.

%\end{CJK}

\end{document}